\documentclass[
reprint,
showpacs,
amsmath,amssymb,aps,
pra,
]{revtex4-1}

\usepackage{graphicx}
\usepackage{dcolumn}
\usepackage{bm}
\usepackage{hyperref}
\usepackage[mathlines]{lineno}
\usepackage{xcolor}

\begin{document}

\preprint{}

\title{Direct temporal mode measurement for the characterization of temporally multiplexed high dimensional quantum entanglement in continuous variables}

\author{Nan Huo$^{1}$}
\author{Yuhong Liu$^{1}$}
\author{Jiamin Li$^{1}$}
\author{Liang Cui$^{1}$}
\author{Xin Chen$^{2}$}
\author{Rithwik Palivela$^{3}$}
\author{Tianqi Xie$^{1}$}
\author{Xiaoying Li$^{1,*}$}
\author{Z. Y. Ou$^{1, 2,\dag}$}
\affiliation{%
$^{1}$College of Precision Instrument and Opto-Electronics Engineering, Key Laboratory of
Opto-Electronics Information Technology, Ministry of Education, Tianjin University,
Tianjin 300072, P. R. China\\
$^{2}$Department of Physics, Indiana University-Purdue University Indianapolis, Indianapolis, IN 46202, USA\\
$^{3}$Carmel High School, 520 E. Main Street, Carmel, IN 46033, USA
}%





\begin{abstract}
Field-orthogonal temporal mode analysis of optical fields is recently developed for a new framework of quantum information science. But so far, the exact profiles of the temporal modes are not known, which makes it difficult to achieve mode selection and de-multiplexing.  Here, we report a novel method that measures directly the exact form of the temporal modes. This in turn enables us to make mode-orthogonal homodyne detection with mode-matched local oscillators. We apply the method to a pulse-pumped, specially engineered fiber parametric amplifier and demonstrate temporally multiplexed multi-dimensional quantum entanglement of continuous variables in telecom wavelength. The temporal mode characterization technique can be generalized to other pulse-excited systems to find their eigen modes for multiplexing in temporal domain.
\end{abstract}


\maketitle

Any electromagnetic field, no matter what state (quantum or classical) it is in, is first characterized by its modes, which are a special class of solutions to the Maxwell equation \cite{lamb}. Modes of the field become especially important when quantum fields are involved because quantum interference requires indistinguishability of photons whereas modes are distinct features of photons. While spatial modes are easily defined through boundary conditions as in optical cavity and waveguide systems, temporal modes are less concerned because they are usually dealt with through spectral analysis in frequency domain.

On the other hand, it was discovered recently that field-orthogonal temporal modes of electromagnetic fields form a new framework for quantum information \cite{raymer}.  Temporal mode analysis offers a straightforward way with intrinsically decoupled modes \cite{sil,lvo,guo} in describing pulse-pumped parametric processes. Parametric processes, ever since first proposed by Yuen \cite{yuen}, have been by far the most common  processes to generate experimentally a variety of quantum states of light \cite{burn,ou,shih,kwiat,sq,ou93,kumar} and are widely used in quantum information and communication \cite{fab17,moran}, quantum simulation \cite{luchao}, and quantum metrology for precision measurement \cite{hud14}.
It should be pointed out that  temporal mode analysis was performed on a pulse-pumped parametric down-conversion of a femtosecond-frequency comb in an optical cavity  with a complicated quantum wavelength multiplexing method \cite{fab14,sch14}, which indirectly revealed the eigen-temporal mode structure as the super modes. Temporal mode functions of photons in spontaneous parametric processes were also obtained indirectly by making the singular-value decomposition of the joint spectral function that can be measured directly \cite{smith}. But as we shall see later, this cannot be applied to high gain parametric processes where quantum entanglement is exhibited in the continuous variables. So, temporal modes of a system has never been directly measured so far.

Moreover, temporal modes are not easily separated even though quantum pulse gates \cite{sil11,sil14,raymer14,raymer18} through nonlinear interaction processes are recently invented to distinguish them with some success. The consequence is that the contributions from different temporal modes add, leading to some detrimental effects  such as extra noise due to out-of-sync phases for different modes in optimum quantum noise reduction \cite{guo}.
On the other hand, homodyne detection can also select out and distinguish the contributions from different temporal modes by a properly matched local oscillator (LO) \cite{guo}. But this requires the knowledge of the exact forms of the temporal modes in order to have a matched LO engineered.

In this paper, we use a feedback-iteration method with a trial seed pulse to obtain and eventually measure the exact forms of the temporal modes of the two correlated fields generated from a pulse-pumped single-pass broadband fiber parametric amplifier. We then measure the quantum correlations between the signal and idler fields by performing homodyne detection with LOs engineered to match the specific temporal modes. We observe quantum entanglement in three pairs of mutually orthogonal temporal modes and confirm the independence between different pairs by quantum measurement.

\begin{figure}
\centering
\includegraphics[width=\linewidth]{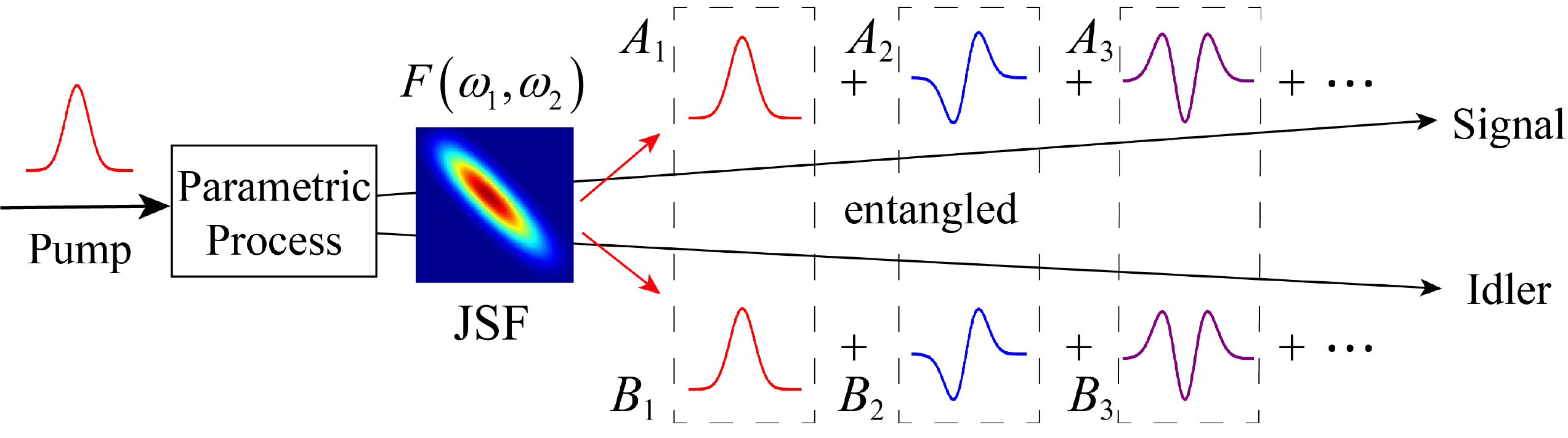}
\caption{Conceptual diagram for the entangled states in various temporal modes ($A_k,B_k$) from a pulse-pumped parametric process. JSF: joint spectral function.
}
\label{fig:ill}
\end{figure}

\noindent{\bf Theoretical background} Pulse-pumped parametric processes generate two fields dubbed ``signal" and ``idler", which are quantum mechanically entangled. The spectral profiles of the entangled fields are extremely complicated because of the dispersion-dependent phase-matching of the nonlinear medium and the wide spectrum of the pump field. However, this complicated system can always be thought of as superposition of its eigen-modes whose temporal/spectral profiles do not change by the amplifier, as shown in Fig.\ref{fig:ill}. That is, there exists an independent set of pairwise modes $\{\hat A_k,\hat B_k\} (k=1,2,...)$ for the signal and idler fields that are related by \cite{sil,lvo,guo}
\begin{eqnarray}\label{in-out}
\hat A^{out}_k &=& \hat A^{in}_k \cosh G_k +  \hat B^{in\dag}_k \sinh G_k\cr
\hat B^{out}_k &=& \hat B^{in}_k \cosh G_k +  \hat A^{in\dag}_k \sinh G_k,
\end{eqnarray}
where $\hat A_k \equiv \int d\omega\psi_k(\omega)\hat a_s(\omega)$, $\hat B_k \equiv \int d\omega\varphi_k(\omega)\hat a_i(\omega)$ are the annihilation operators for the $k$-th modes of the signal and idler fields with respective eigen-temporal profiles of $f_k(\tau)\equiv \int d\omega \psi_k(\omega) e^{-i\omega\tau}, g_k(\tau)\equiv \int d\omega \varphi_k(\omega) e^{-i\omega\tau}$, satisfying the ortho-normal relations
\begin{eqnarray}\label{ornorm}
\int d\omega\psi^*_k(\omega) \psi_{k'}(\omega) = \delta_{kk'} = \int d\omega\varphi_k^*(\omega) \varphi_{k'}(\omega).~~~~
\end{eqnarray}
These eigen-modes are exactly the super modes studied by Roslund {\it et al} \cite{fab14}, which are pairwise entangled and form a multi-dimensional quantum entangled states.

At relatively low pumping power so that $|G_k|\ll 1$, the eigen-functions $\{\psi_k(\omega),\varphi_k(\omega)\}$ can be obtained by singular value decomposition (SVD) method from the joint spectral function (JSF)
\begin{eqnarray}\label{JSF}
F(\omega_1,\omega_2)= G \sum_k r_k \psi_k(\omega_1) \varphi_k(\omega_2),
\end{eqnarray}
which is defined via the interaction Hamiltonian \cite{sil,ou-multi}
\begin{eqnarray}\label{H}
\frac{1}{i\hbar}\int dt \hat H   = \int d \omega_1 d\omega_2F(\omega_1,\omega_2)\hat a_s ^{\dag}(\omega_1)\hat a_i^{\dag}(\omega_2)+ h.c.~~~~
\end{eqnarray}
Here, $\{r_k\}\ge 0$ are the mode numbers satisfying the normalization relation $\sum_k r_k^2=1$. $G>0 $ is a parameter proportional to the peak amplitudes of the pump fields, nonlinear coefficient and the nonlinear medium length. Then, we have $G_k=r_k G$ with $r_k$ independent of $G$.

At high pump power when stimulated emission dominates, Eq.(\ref{in-out}) still holds but $\{r_k\}$ and $\{\psi_k(\omega),\varphi_k(\omega)\}$ now depend on the pump parameter $G$ \cite{sipe,sam}.

\begin{figure}
\centering
\includegraphics[width=\linewidth]{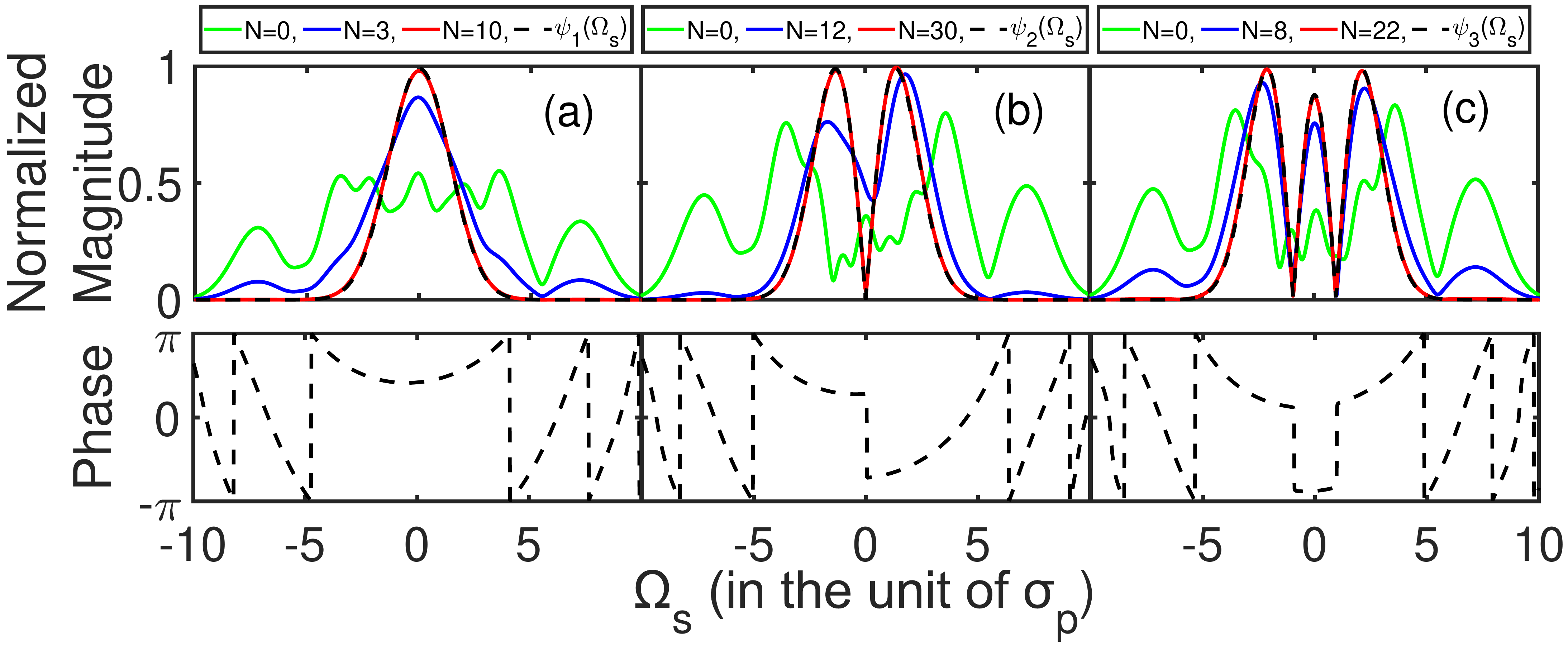}
\caption{Simulated output spectral amplitudes $|\psi_k(\omega)|$ (top) and converged phases (bottom) in the unit of pump bandwidth $\sigma_p$ for (a) $k=$1, (b) $k=$2, (c) $k=$3. The green curves are the input spectral functions while the blue and red curves are intermediate outputs after the iteration steps indicated in the legends. The dashed curves are the final outputs.
}
\label{fig1:simu}
\end{figure}

\begin{figure*}[htbp]
\centering
\includegraphics[width=15cm]{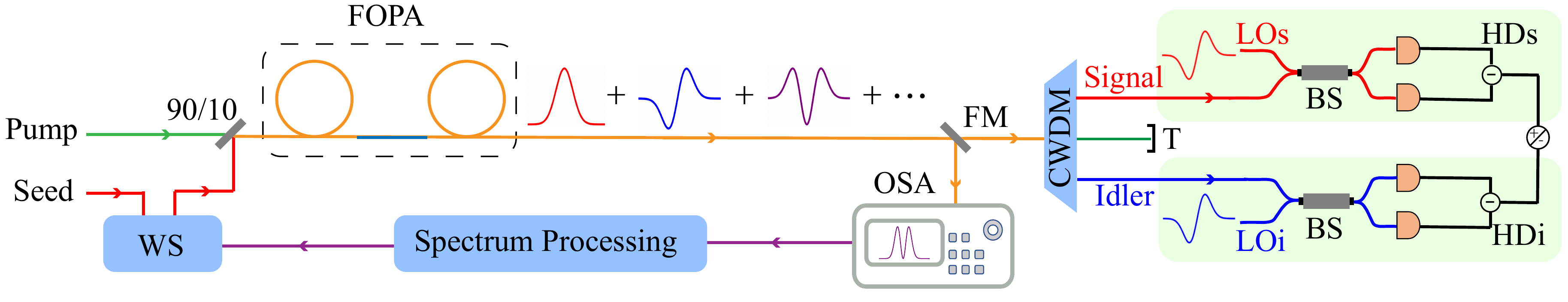}
\caption{Experimental setup. WS, wave shaper; FOPA, fiber optical parametric amplifier; OSA, optical spectrum analyzer; FM, flip mirror; CWDM, coarse wavelength division multiplexer; BS, beam splitter; T, terminator; HD, homodyne detection.
}
\label{fig:setup}
\end{figure*}

So far, mode functions $\{\psi_k(\omega_1),\varphi_k(\omega_2)\}$ are only obtained in simulations but have never been measured directly. In the following, we will describe a method to directly measure these mode functions experimentally.
\vskip 0.1in

\noindent {\bf Temporal modes determination}
Our procedure to find the mode functions $\psi_k(\omega),\varphi_k(\omega)$ is based on Eq.(\ref{in-out}). We inject a seed into the signal field and observe its output. This is somewhat similar to the method of stimulated emission tomography \cite{sipe2,lor}. But here, after the measurement of the output, we feed the result back to modify the input seed and iterate the process. This part is similar to the adaptive method of Polycarpou {\it et al} \cite{adap}. To see what this leads to, we consider a coherent pulse of spectral shape $\alpha_0(\omega)$ injected into field $A$. Because of the orthonormality in Eq.(\ref{ornorm}), we can expand it as
\begin{eqnarray}\label{in}
\alpha_0(\omega)= \sum_k \xi_k \psi_k(\omega)
\end{eqnarray}
with $\xi_k = \int d\omega \psi_k^*(\omega)\alpha_0(\omega)$. Using Eq.(\ref{in-out}) and assuming $|\xi_k|^2\gg1$ to ignore spontaneous emission, we find
\begin{eqnarray}\label{out}
\alpha^{out}(\omega)= \sum_k \xi_k \cosh G_k \psi_k(\omega).
\end{eqnarray}
So, each mode is amplified but with different gain. Now let us exploit this gain difference: we can measure the output spectral shape and then program a new input field with a wave shaper according to the measured shape. To keep the input power low, we can attenuate the output by a factor, say $(\cosh G_1)^{-1}$, so that the new input becomes
\begin{eqnarray}\label{out2}
\alpha_1(\omega)= \sum_k \xi_k (\cosh G_k/\cosh G_1) \psi_k(\omega).
\end{eqnarray}
Since $G_k$'s are different for different $k$, let us arrange mode order: $G_1>G_2>...$ and $\cosh G_k/\cosh G_1 < 1$ for all except $k=1$. We iterate the procedure $N$ times and the field after $N$ iterations becomes
\begin{eqnarray}\label{out3}
\alpha_N(\omega)= \sum_k \xi_k (\cosh G_k/\cosh G_1)^N \psi_k(\omega).
\end{eqnarray}
With $N$ large enough, $(\cosh G_k/\cosh G_1)^N\rightarrow 0$ for $k\ne 1$ and we are left with only the first mode: $\alpha_N(\omega)\propto \psi_1(\omega)$.

To obtain the mode function for $k=2$, we need to have an input field that is orthogonal to $\psi_1(\omega)$, that is, $\xi_1=0$. To achieve this, we use the Gram-Schmidt process: with $\psi_1(\omega)$ known, we set the input as $\alpha'(\omega) = \alpha(\omega) - \xi_1 \psi_1(\omega)$, which has $\xi_1'=0$. Then the dominating mode is $k=2$. We perform orthogonalization after each iteration to ensure $\xi_1'=0$. Subsequent modes can be obtained in a similar way but with the orthogonalization changed to $\alpha'(\omega) = \alpha(\omega) - \sum_{i=1}^{k-1}\xi_i \psi_i(\omega)$ for mode $k$.

In order to demonstrate the validity of the method, we run some simulations based on Eq.(\ref{out3}) for the JSF given in Ref.\cite{guo} but with a chirped pump phase of $e^{i\Omega^2/2\sigma_p^2}$ and set $G=2.5$. The results are shown in Fig.\ref{fig1:simu} for the first three modes. The green and dashed curves are the initial input and the final output spectral functions, respectively. The blue and red curves are for the intermediate steps with the step numbers shown in the legends. Only the final converged shapes are shown for phase (bottom).

\noindent {\bf Experimental Procedures and Results}
The experimental setup is shown in Fig.\ref{fig:setup}, in which the pulse-pumped fiber optical parametric amplifier (FOPA) consists of two dispersion-shifted fibers (DSFs) and a single-mode fiber (SMF), which, through a quantum interference effect, modifies the JSF so that it is well-behaved for the iteration method to converge \cite{su} (see Supplementary Materials for details). The pump and the seed, with their path lengths carefully balanced through a delay line (not shown), are combined with a 90/10 beam splitter and simultaneously launched into the FOPA. The output of FOPA is either measured by an optical spectral analyzer (OSA) to determine the spectral profile or separated by coarse wavelength division multiplexer (CWDM) for quantum measurement by homodyne detection.

\begin{figure}[htbp]
\centering
\includegraphics[width=\linewidth]{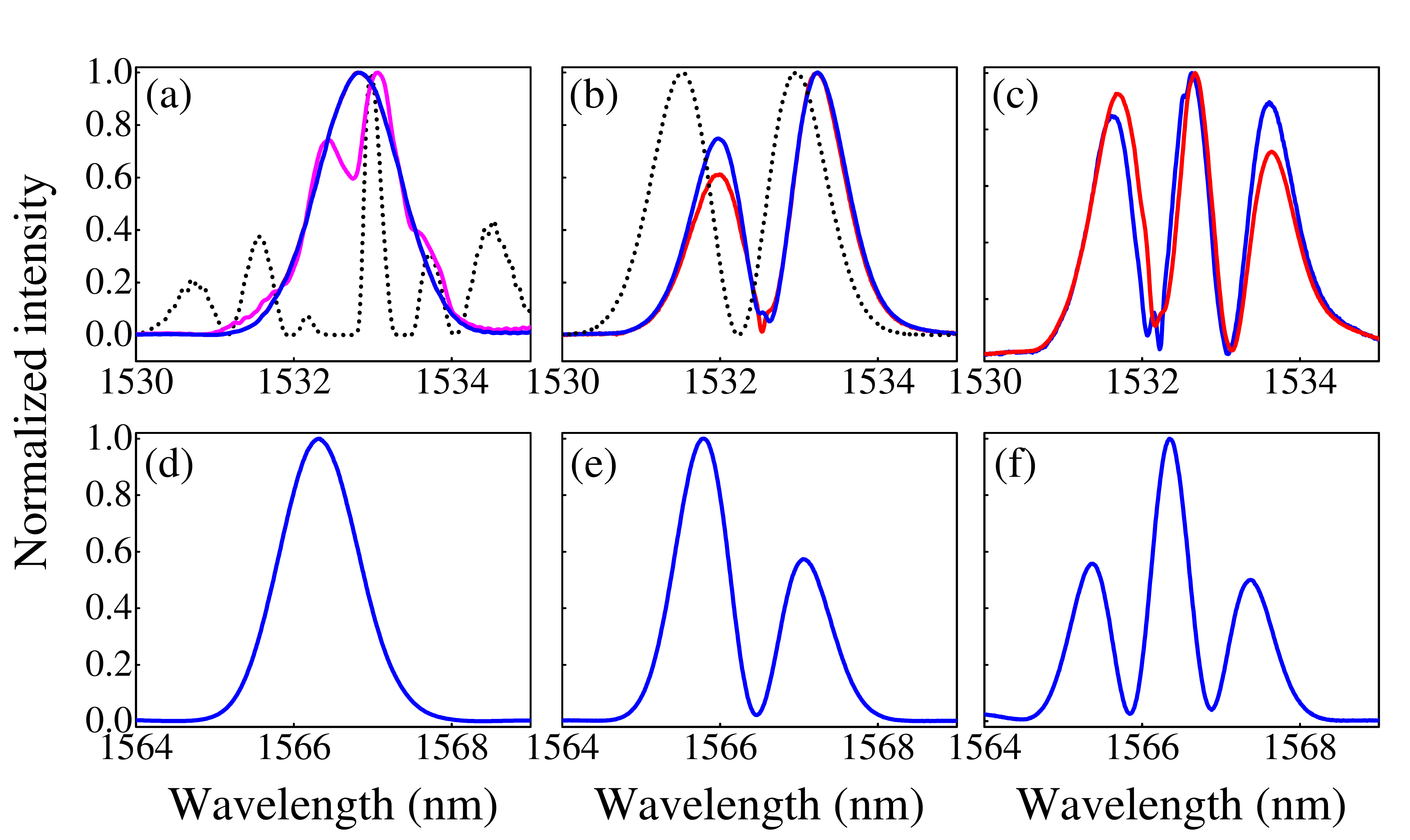}
\caption{Measured spectral intensity $|\psi_k(\omega)|^2$ for $k=$ (a) 1, (b) 2, (c) 3,  and those ((d),(e),(f)) for the corresponding idler field. The dotted lines are the initially injected seed (not shown for (c) due to crowdedness). The pink curve in (a) is the output after two iterations. The blue lines are the output signal and the red lines are the feedback to the input (red is covered by blue in (a)).
}
\label{fig:measured}
\end{figure}

We first determine directly the temporal mode profiles of the fiber parametric amplifier by the feedback-iteration method described previously. For this, we use the recorded spectrum of the signal field by an optical spectral analyzer (OSA) to reshape the input seed with a wave shaper (WS). Although an OSA only measures the spectral intensity, here, in the first order approximation, we assume that there is no dispersion in the phases of the mode functions except a jump of $\pi$ at zeros for higher order modes.  Such an assumption is valid because the spectral phases are relatively flat within the spectral width ($\sim$ 3.5 nm) of the specially engineered source (see Ref. [31] and Supplementary Materials). So we implement a sign change for the high order mode cases whenever the spectral intensity goes to zero.  After a number of iterations ($\sim 6-8$  depending on the shape of the initial injection), a steady shape is reached, which corresponds to one of the eigen temporal modes from the parametric amplifier. We follow the steps described previously to find other eigen temporal modes. The blue curves in Fig.\ref{fig:measured} are the converged spectral intensity of the first three temporal modes (a,b,c) together with those for the corresponding idler field (d,e,f). The curves are normalized to the maximum values. The dotted lines are the initially injected seed (only for (a) and (b)).  The pink curve in Fig.\ref{fig:measured}(a) is the output after only two iterations, showing fast convergence of the iteration. For the higher order modes ($k=2,3$), there is a slight difference between the feedback input (red) and the output (blue). This is caused by the non-uniform spectral response of the detector as well as dispersion in phase of the higher order modes.

\begin{figure}[htbp]
\centering
\includegraphics[width=\linewidth]{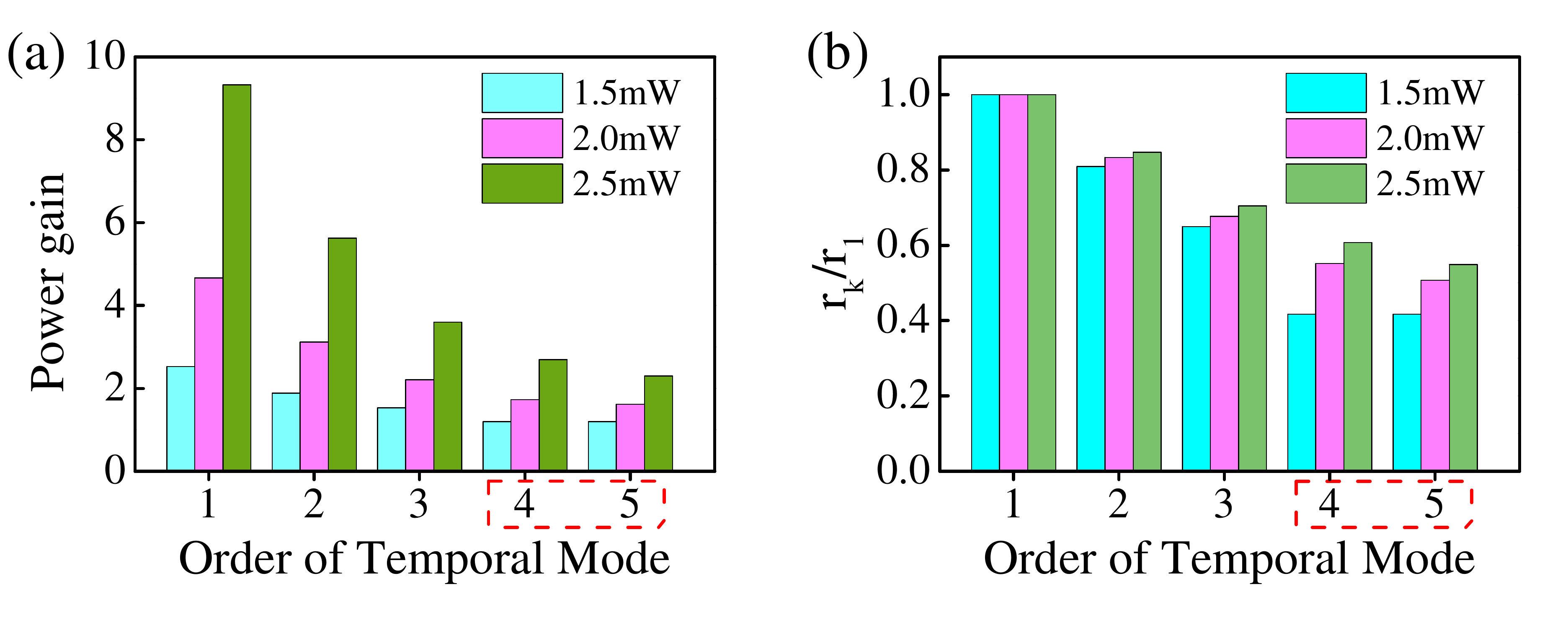}
\caption{(a) Power gain for the first five modes under different pump power. (b) Extracted mode numbers. The dashed red boxes on order numbers 4 and 5 indicate less reliable results due to unstable output.}
\label{fig:gain}
\end{figure}

The temporal mode structure is characterized by the distribution of the $G_k$ values, which can be obtained from the power gain $\cosh^2 G_k$ for each mode. The measured power gains for the first five modes under different pump powers are shown in Fig.\ref{fig:gain}(a) with the extracted $r_k/r_1 (\equiv G_k/G_1)$ values shown in  Fig.\ref{fig:gain}(b). The dashed red boxes on order numbers 4 and 5 in the figure indicate that the output is not stable. This is because the bandwidths of the higher orders are too broad and run outside the range of the well-behaved JSF and into the next region of the JSF (see Supplementary Materials on JSF). As can be seen from Fig.\ref{fig:gain}(b), the mode parameters $\{r_k\}$ change with pump power increase and the changes become more prominent for the high orders. Furthermore, we also observe some significant changes in the converged mode profiles (Fig.\ref{fig:measured}) as the pump power changes. This is in support of the theory in Refs. \cite{sipe,sam} that mode structure changes with the pump power. But the mode changes may also be caused by other nonlinear effects in fibers such as self-phase modulation, which occur at high pump power.

Once the temporal mode profiles are determined, we can perform homodyne detection with local oscillators (LOs or LOi) tailored to match the specific temporal mode of our interest. Since temporal modes are orthogonal to each other, when the LO is matched to a specific mode, there is no contribution from other modes for the homodyne detection \cite{guo}. So we can use homodyne detection to select the mode of our interest. With this, we first check the correlation between different modes by measuring the covariance matrix $C_{mn} \equiv \langle \Delta \hat O_m\Delta \hat O_n\rangle/\sqrt{\langle \Delta^2 \hat O_m\rangle\langle\Delta^2 \hat O_n\rangle}$ where indices $m, n$ denote different modes and $\hat O $ is either the amplitude quadrature $\hat X \equiv \hat a+\hat a^{\dag}$ or the phase quadrature $\hat Y\equiv (\hat a - \hat a^{\dag})/i$ when properly locking the phases between LOs/LOi and signal/idler beams to either 0 or $\pi/2$ (see Supplementary Materials for the details). Figure \ref{fig:cov} shows the results of either amplitude (Fig.\ref{fig:cov}(a)) or phase (Fig.\ref{fig:cov}(b)) quadrature for six modes of $\{s1,s2,s3,i3,i2,i1\}$ with $sk,ik (k=1,2,3)$ denoting the k-th order conjugate modes of signal and idler fields (For exact numerical values, see Supplementary Materials). We only measure up to 3 orders because higher orders are not stable (see Fig.\ref{fig:gain}). In Fig.\ref{fig:cov}, we take out the diagonal elements which all equal to 1. The anti-diagonal elements correspond to $C_{s1i1}, C_{s2i2}, C_{s3i3}$ whose non-zero values show the strong pairwise correlation (amplitude $\hat X$) or anti-correlation (phase $\hat Y$) between the corresponding signal and idler modes of the same order  whereas the other off-diagonal elements are near zero, indicating the total independence between different orders of the temporal modes and confirming the orthogonality of the modes.

\begin{figure}[htbp]
\centering
\includegraphics[width=\linewidth]{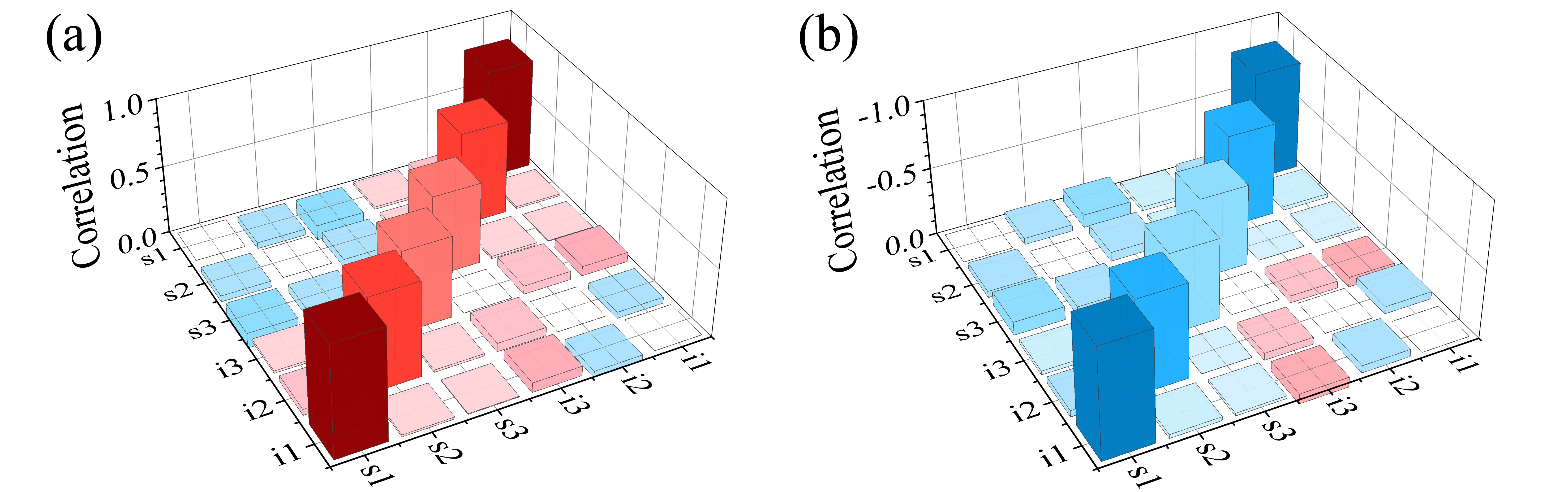}
\caption{Covariance matrix elements for (a) amplitude quadrature (X) and (b) phase quadrature (Y). The trivial diagonal elements of $C_{ii}=1$ are taken out while the non-zero anti-diagonal elements shows strong pair-wise correlation between the corresponding modes of signal and idler beams. Other near zero off-diagonal elements indicate total independence between different modes. See Supplementary Materials for exact numerical values.}
\label{fig:cov}
\end{figure}

Next, we check the quantum entanglement between the signal and idler fields by measuring  $\langle \Delta^2(\hat X_{sk}-\hat X_{ik})\rangle$ and  $\langle \Delta^2(\hat Y_{sk}+\hat Y_{ik})\rangle$ ($k=1,2,3$) for the $k$-th modes and verifying the inseparability criterion of entanglement: $I_k\equiv \langle \Delta^2(\hat X_{sk}-\hat X_{ik})\rangle/\langle \Delta^2(\hat X_{sk}-\hat X_{ik})\rangle_u+\langle \Delta^2(\hat Y_{sk}+\hat Y_{ik})\rangle/\langle \Delta^2(\hat Y_{sk}+\hat Y_{ik})\rangle_u < I_u = 2$, with the subscript $u$ denoting the unentangled vacuum case \cite{ent}. At pump power of about 1.3 mW, resulting in power gains of the FOPA of 2.1, 1.5 and 1.3 for the first three modes, the measured values of $I_k/I_u (k=1,2,3)$ are obtained in log-scale as $-2.56$ dB, $-1.50$ dB and $-1.17$ dB for the first three order modes, respectively.

Taking the total detection efficiency into consideration, we obtain the corrected value of $I/I_u$ as $-3.70$ dB, $-2.00$ dB and $-1.60$ dB or $I=0.85,1.26, 1.38 < 2$ for the first three modes, respectively. This shows pairwise entanglement between the signal and idler modes of the same order. The worsening for the higher orders is mostly because of the decreasing gains for the higher order modes but is also partly because of the mode mis-matching due to inaccurate mode function measurement.

In summary, we show theoretically and demonstrate experimentally a method that directly determines the temporal/spectral profiles of the eigen-modes for
the signal and idler fields generated from a fiber-based
parametric amplifier pumped by a short pulse. We further show experimentally that they are pairwise entangled.

In our proof-of-principle experiment, we do not measure phase part of the pulses because we assume  constant phase across the spectrum, which is fulfilled by the specially engineered source. Our experimental approach thus cannot measure any spectral phase variations and won't work for sources with large phase changes. Our simulations, on the other hand, demonstrate the general applicability of the method once intensity and phase are both considered (Fig.\ref{fig1:simu}). So, for the general case, a full measurement in both intensity and phase is necessary.  This can be done via pulse characterization techniques \cite{spider} such as FROG \cite{treb}, the interferometric method \cite{wam} and cross-correlation FROG \cite{xfrog}.

We also demonstrate that the mode structure depends on the gain of the parametric amplifier. But the method described here is only suitable for the high gain parametric amplifier, which gives rise to quantum entanglement in continuous variables. For the low gain case, which produces a two-photon entangled state, the method fails because the gain is near unity for all modes. Nevertheless, the feedback-iteration idea in the current method can still be applied to the low gain case by involving the stimulated emission in the idler field. The detail is presented in another paper  \cite{xchen}.

The technique can be generalized to other  pulse-pumped systems such as frequency conversion process or other degrees of freedom such as spatial modes to find the eigen-modes of the system. So, the potential applications of the technique is not limited only to quantum optics but can be applied to classical systems as well.

\begin{acknowledgments}
We would like to thank S. Lemieux and R. W. Boyd for helpful discussion.
This work was supported in part by National Natural Science Foundation of China (91836302, 91736105, 11527808), the National Key Research and Development Program of China (2016YFA0301403), and in part by US National Science Foundation (Grant No. 1806425).
\end{acknowledgments}




\begin{thebibliography}{10}
\newcommand{\enquote}[1]{``#1''}

\bibitem{lamb} W. E. Lamb, Appl. Phys. B {\bf 60}, 77 (1995).

\bibitem{raymer} B. Brecht, Dileep V. Reddy, C. Silberhorn, and M. G. Raymer, Phys. Rev. X {\bf 5}, 041017 (2015).

\bibitem{sil} A. Christ, K. Laiho, A. Eckstein, K. N. Cassemiro, and C. Silberhorn, New J. Phys. {\bf 13}, 033027 (2011).

\bibitem{lvo} W. Wasilewski, A. I. Lvovsky, K. Banaszek, and C. Radzewicz, Phys. Rev. A {\bf 73}, 063819 (2006).

\bibitem{guo} Xueshi Guo, Nannan Liu, Xiaoying Li, and Z. Y. Ou, Opt. Express \textbf{23}, 029369 (2015).


\bibitem{yuen} H. P. Yuen, Phys. Rev. A {\bf 13}, 2226 (1976).

\bibitem{burn} D. C. Burnham and D. L. Weinberg, Phys. Rev. Lett. {\bf 25}, 84 (1970).

\bibitem{sq} R. E. Slusher, L. W. Hollberg, B. Yurke, J. C. Mertz, and J. F. Valley, Phys. Rev. Lett. {\bf 55}, 2409 (1985).

\bibitem{ou} Z. Y. Ou and L. Mandel, Phys. Rev. Lett. {\bf 61}, 50 (1988).

\bibitem{shih} Y. H. Shih and C. O. Alley, Phys. Rev. Lett. {\bf 61}, 2921 (1988).

\bibitem{kwiat} P. G. Kwiat, K. Mattle, H. Weinfurter, A. Zeilinger, A. V. Sergienko, and Y.
Shih, Phys. Rev. Lett. {\bf 75}, 4337 (1995).

\bibitem{ou93} Z. Y. Ou,   S. F. Pereira, H. J. Kimble, and K. C. Peng, Phys. Rev. Lett. {\bf 68}, 3663 (1992).

\bibitem{kumar} O. Ayt\"ur  and P. Kumar, Phys. Rev. Lett.
{\bf 65}, 1551 (1990).

\bibitem{fab17} Y. Cai,  J. Roslund, G. Ferrini, F. Arzani, X. Xu, C. Fabre and  N. Treps, Nat. Comm. {\bf 8}, 15645 (2017).

\bibitem{moran} C. Reimer, S. Sciara, P. Roztocki, M. Islam, L. Romero Cort\'es,
Y. Zhang, B. Fischer, S. Loranger, R. Kashyap, A. Cino,
S. T. Chu, B. E. Little, D. J. Moss, L. Caspani, W. J. Munro, J. Azana,
M. Kues, and R. Morandotti, Nat. Phys. {\bf 15}, 148 (2019).

\bibitem{luchao} H. Wang, Y. He, Y.-H. Li, Z.-E. Su, B. Li, H.-L. Huang, X. Ding, M.-C. Chen, C. Liu, J. Qin, J.-P. Li, Y.-M.
He, C. Schneider, M. Kamp, C.-Z. Peng, S. H\"ofling, C.-Y. Lu, and J.-W. Pan, Nat. Photonics {\bf 11}, 361 (2017).

\bibitem{hud14} F. Hudelist, J. Kong, C. Liu, J. Jing, Z. Y. Ou, and W. Zhang, Nature Commun. {\bf 5}, 3049 (2014).

\bibitem{fab14} Jonathan Roslund, Renne Medeiros de Araujo, Shifeng Jiang, Claude Fabre, and Nicolas Treps, Nat. Photonics \textbf{8}, 109 (2014).

\bibitem{sch14} R. Schmeissner, J. Roslund, C. Fabre, and N. Treps, Phys. Rev. Lett. {\bf 113}, 263906 (2014).

\bibitem{smith} Alex O. C. Davis, Val\'erian Thiel, and Brian J. Smith, arXiv:1809.03727 (2018).

\bibitem{sil11} A. Eckstein, B. Brecht, and C. Silberhorn, Opt. Express {\bf 19}, 13770 (2011).

\bibitem{sil14} B. Brecht, A. Eckstein, R. Ricken, V. Quiring, H. Suche, L. Sansoni, and C. Silberhorn, Phys. Rev. A {\bf 90}, 030302(R) (2014).

\bibitem{raymer14} D. V. Reddy, M. G. Raymer, and C. J. McKinstrie, Opt. Lett. {\bf 39}, 2924 (2014).

\bibitem{raymer18} D. V. Reddy and M. G. Raymer, Optica \textbf{5}, 423 (2018).

\bibitem{ou-multi} Zhe-Yu Jeff Ou, \textit{Quantum Multi-Photon Interference}, (Springer, New York, 2007).

\bibitem{sipe} N. Quesada and J. E. Sipe, Phys. Rev. A {\bf 90}, 063840 (2014).

\bibitem{sam} P. R. Sharapova, G. Frascella, M. Riabinin, A. M. P\'erez, O. V. Tikhonova,
S. Lemieux, R. W. Boyd, G. Leuchs, and M. V. Chekhova, arXiv:1905.10109 (2019).

\bibitem{sipe2} M. Liscidini and J. E. Sipe, Phys. Rev. Lett. {\bf 111}, 193602 (2013).

\bibitem{lor} B. Fang, O. Cohen, M. Liscidini, J. E. Sipe, and V. O. Lorenz, Optica \textbf{1}, 281 (2014).

\bibitem{adap} C. Polycarpou, K. N. Cassemiro, G. Venturi, A. Zavatta, and M. Bellini, Phys. Rev. Lett. {\bf 109}, 053602 (2012).

\bibitem{su} J. Su, L. Cui, J. Li, Y. Liu, Xiaoying Li, and Z. Y. Ou, Opt. Express \textbf{27}, 020479 (2019).



\bibitem{ent} L.-M. Duan, G. Giedke, J. I. Cirac, and P. Zoller, Phys. Rev. Lett. {\bf 84}, 2722 (2000).

\bibitem{spider} I. A. Walmsley and C. Dorrer, Advances in Optics and Photonics {\bf  1}, 308 (2009).

\bibitem{treb} Rick Trebino, Kenneth W. DeLong, David N. Fittinghoff, John N. Sweetser,
Marco A. Krumb\"ugel, Bruce A. Richman, and Daniel J. Kane, Rev. Sci. Instrum. {\bf  68}, 3277 (1997).

\bibitem{wam} D. N. Fittinghoff, J.L. Bowie, J.N. Sweetser, R. T. Jennings, M. A. Krumbuegel, K. W. DeLong, R. Trebino, and I. A. Walmsley, Opt. Lett. {\bf 21}, 884 (1996).

\bibitem{xfrog} Jing-yuan Zhang, Aparna Prasad Shreenath, Mark Kimmel, Erik Zeek, Rick Trebino and Stephan Link, Opt. Express, {\bf 6}, 601 (2003).

\bibitem{xchen} X. Chen, Xiaoying Li, and Z. Y. Ou, arXiv:1910.09720 (2019), to appear in Phys. Rev. A (2020).

\bibitem{Su-arch} J. Su, L. Cui, J. Li, Y. Liu, X. Li, and Z. Y. Ou, arXiv:1811.07646 (2018).



\bibitem{renyongoe05} R. Tang, J. Lasri, P. S. Devgan, V. Grigoryan, and P. Kumar, Opt. Express {\bf 13}, 10483 (2005).


\bibitem{yuhong} Y, Liu, J. Li, L. Cui, N. Huo, S. M. Assad, Xiaoying Li, and Z. Y. Ou, Opt. Express {\bf 26}, 27705 (2018).


%
%
%
%
%






\end{thebibliography}

\vfil \eject

\vskip 0.2in

{\hskip 0.6 in  \bf Supplementary Materials}

\begin{figure}[htb]
	\centering
	\includegraphics[width=8cm]{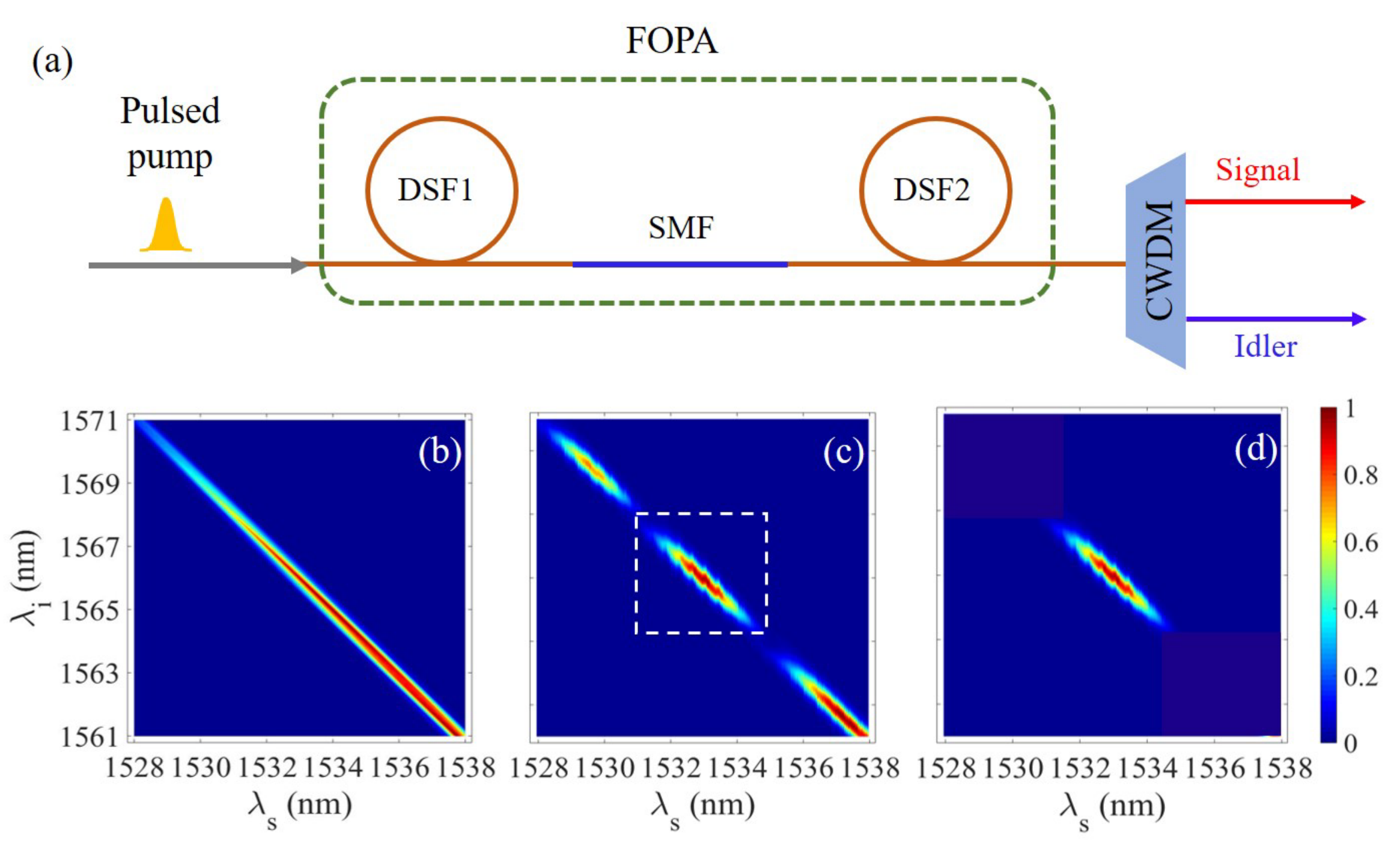}
	\caption{(a) The scheme of specially engineered fiber optical parametric amplifier (FOPA), formed
		by two identical dispersion shifted fibers (DSF) with a standard single-mode fiber (SMF)
		in between.  The joint spectral intensity $|F(\omega_1,\omega_2)|^2$ of signal and idler field for a single DSF fiber (b), for two DSF fibers with a SMF in between (c), and  after CWDM (d). }
	\label{SUI}
\end{figure}

\vskip 0.2in
\noindent{\bf I. Specially Engineered Fiber Optical Parametric Amplifier} -- The fiber optical parametric amplifier (FOPA) in Fig.3 in the main text is formed by a two-stage nonlinear interferometer (NLI), which functions as an amplifier with specially engineered joint spectral function (JSF) \cite{Su-arch}. The FOPA consists of two identical
DSFs with a piece of standard single mode fiber (SMF) in between, as shown in Fig.\ref{SUI}(a). The lengths of each DSF and SMF are 150 m and 3.4 m, respectively. The zero group velocity dispersion (GVD)
wavelength and GVD slope of each DSF are 1548.5 nm and 0.075 ps/$km$/$nm^{2}$, respectively, and the nonlinear coefficient of each DSF is
about 2 $(W\cdot Km)^{-1}$. The GVD coefficient of the SMF is 17
ps/$(km\cdot nm)$ in the vicinity of 1550 nm band. The pulsed pump of FOPA is centered at 1549.32 nm to ensure that the phase matching of the four-wave mixing (FWM) parametric process is satisfied in DSFs. Note that the DSFs serve as nonlinear media of FWM while the standard
SMF, in which the phase matching of FWM is not satisfied, functions as the linear dispersive medium to modify the JSF of the FOPA.

The reason that we use this specially engineered FOPA is because the JSF directly out of a single piece of DSF has a wide spectrum, as shown in Fig.\ref{SUI}(b),  that overlaps with that of the pump and thus is influenced by the pump. The JSF of our FOPA is shown in Fig.\ref{SUI}(c). One sees that the JSF is split into islands due to the active filtering mechanism originated from interference between the two FWM processes in DSFs \cite{Su-arch}. In this experiment, because the overlap between two adjacent islands decreases with the bandwidth of pump \cite{Su-arch,renyongoe05}, in order to better isolate the island from each other, the spectrum of the pump, which is originated from an ultra-fast mode-locked fiber laser of repetition rate of 50 MHz, is carved by a narrow band filter with a full width at half maximum (FWHM) of 0.28 nm. The chirp of pump is negligible due to smallness. Moreover, the length of SMF between two DSFs in FOPA is set to 3.4 m to ensure the island of our interest is within the spectrum ($1571\pm 8$ nm for signal and $1531\pm 8$ nm for idler) at which the filter for efficiently isolating strong pump and transmitting signal and idler fields is available. When the output of FOPA propagates through a coarse wavelength division multiplexer (CWDM), the signal and idler beams are then separated from the strong pump. The CWDM has two channels centered at 1531 and 1570 nm,
respectively. For each channel, the isolation to the pump is greater
than 60 dB, and the one-dB bandwidth for each CWDM channel is 16
nm. Hence, Fig.\ref{SUI}(d) can present the JSF of signal and idler fields at the output of CWDM, which corresponds to the island highlighted by the dashed box in Fig.\ref{SUI}(c).

\vskip 0.2in

\noindent {\bf II. Phase Locking} -- We achieve phase locking of the local oscillators by passing the injected seed sequentially through a phase modulator (PM) and an amplitude modulator (AM) \cite{yuhong}. In this case, both the modulated signals of the PM and AM are transferred to the amplified signal and idler beams to produce error signals for locking. When the relative phases are locked to 0 by exploiting the sinusoidal modulation signal of the PM at 2.5 MHz, we are able to measure the noise of quadrature amplitudes  $\hat X_s$ and $\hat X_i$; when the relative phases are locked to $\pi/2$  by using the sinusoidal modulation signal of the AM at 1.875 MHz, we are able to measure the noise of quadrature phase  $\hat Y_s$ and $\hat Y_i$.
\vskip 0.2in
\noindent {\bf III. Covariance Matrix} -- To obtain the covariance matrix elements $C_{mn}$ defined as
\begin{eqnarray}\label{C-def}
C_{mn} \equiv \frac{\langle \Delta \hat O_m\Delta \hat O_n\rangle}{\sqrt{\langle \Delta^2 \hat O_m\rangle}\sqrt{\langle\Delta^2 \hat O_n\rangle}}
\end{eqnarray}
for $\hat O_m,\hat O_n (O=X, Y, m, n=s1,s2,s3,i3,i2,i1)$,
we first measure $\langle \Delta^2 \hat O_m\rangle$ by direct homodyne detection of $\hat O_m$. $\langle \Delta \hat O_m\Delta \hat O_n\rangle$ can be extracted from the directly measured joint quantity $\langle \Delta^2 (\hat O_m- \hat O_n)\rangle \equiv  \langle \Delta^2 \hat O_m\rangle + \langle\Delta^2 \hat O_n \rangle -2\langle \Delta \hat O_m\Delta \hat O_n\rangle$.

The covariance matrix elements between signal and idler beams such as $C_{s1i1}$ can be measured from joint measurement of $\Delta^2(\hat X_{s1}- \hat X_{i1})$ by homodyne detection on the separated signal and idler beams. But matrix elements such as $C_{s1s2}$ will have to come from measurement on signal beam alone. Since we don't yet have an effective method to separate the temporal modes, we cannot use the method of measuring $C_{s1i1}$. Instead, we consider an LO in a shape of the superposition $\psi_{LO}(\omega) = \alpha_1 \psi_{s1}(\omega) + \alpha_2\psi_{s2}(\omega) (\alpha_j$ is real). It can be shown \cite{guo} that with proper phase of the LO field, the homodyne detection with this LO measures the quantity
\begin{eqnarray}\label{HD-m}
\hat X_m &=& \int d\omega \psi_{LO}^*(\omega)\hat E^{(+)}(\omega) + h.c. \cr
&=& \alpha_1\hat X_{s1} + \alpha_2\hat X_{s2},
\end{eqnarray}
where $\hat X_{sj} \equiv \hat A_{sj} +\hat A_{sj}^{\dag} (j=1,2...)$ with $\hat A_{sj}$ defined in Eq.(1) in the main text as the annihilation operator for mode $sj$. Then we have
\begin{eqnarray}\label{HD-m-D}
\langle \Delta^2 \hat X_m\rangle
&=&\alpha_1^2\langle \Delta^2\hat X_{s1} \rangle + \alpha_2^2 \langle \Delta^2\hat X_{s2}\rangle \cr
&&~~~~+ 2\alpha_1\alpha_2\langle \Delta\hat X_{s1} \Delta \hat X_{s2}\rangle.
\end{eqnarray}
Making three measurement by setting $\alpha_1=1, \alpha_2=0$, or $\alpha_1=0, \alpha_2=1$, or $\alpha_1=1, \alpha_2=1$, respectively, we are able to extract $C_{s1s2}$ even though we cannot separate $s1$ and $s2$ modes.

The values of the elements of the measured covariance matrix with labels $\{1=s1,2=s2,3=s3,4=i3,5=i2,6=i1\}$ are presented  for amplitude ($\hat X$) as
\begin{eqnarray}\label{CX}
&&C^X_{mn} =\cr
&& \left(\begin{matrix}
1.00 &-0.05 &-0.11 &0.02 &0.04 &0.82\\ -0.05 &1.00 &-0.06 &0.03 &0.69 &0.01\\ -0.11 &-0.06 &1.00 &0.60 &0.02 &0.01\\
0.02 &0.03 &0.60 &1.00 &0.06 &0.07 \\ 0.04  &0.69 &0.02 &0.06 &1.00 &-0.05 \\ 0.82  &0.01 &0.01 &0.07 &-0.05 &1.00 \\
\end{matrix}\right),\cr &&
\end{eqnarray}
and for phase ($\hat Y$) as
\begin{eqnarray}\label{CY}
&&C^Y_{mn} =\cr
&& \left(\begin{matrix}
1.00 &-0.05 &-0.10 &-0.03 &-0.05 &-0.81\\ -0.05 &1.00 &-0.07 &-0.02 &-0.69 &-0.03\\ -0.10 &-0.07 &1.00 &-0.60 & 0.00 &-0.02\\
-0.03 &-0.02 &-0.60 &1.00 &0.05 &0.07 \\ -0.05  &-0.69 & 0.00 &0.05 &1.00 &-0.05 \\ -0.81  &-0.03 &-0.02 &0.07 &-0.05 &1.00 \\
\end{matrix}\right).\cr &&
\end{eqnarray}
The error bars for all elements are approximately 0.01. The diagonal elements are trivial value of 1. The anti-diagonal elements are near 1 or -1, indicating strong correlation between the signal and idler fields of the same order of modes $s_m,i_m (m=1,2,3)$. The other off-diagonal elements are close to zero indicating the independence of the modes of different orders.

\end{document}